\documentclass{article}

\PassOptionsToPackage{numbers, sort&compress}{natbib}

\usepackage[preprint]{neurips_2023_ml4ps}
\usepackage{graphicx}       % Required for inserting images
\usepackage[utf8]{inputenc} % allow utf-8 input
\usepackage[T1]{fontenc}    % use 8-bit T1 fonts
\usepackage{hyperref}       % hyperlinks
\usepackage{url}            % simple URL typesetting
\usepackage{booktabs}       % professional-quality tables
\usepackage{amsfonts}       % blackboard math symbols
\usepackage{nicefrac}       % compact symbols for 1/2, etc.
\usepackage{microtype}      % microtypography
\usepackage{caption}        % captionof
\usepackage{amsmath}        % pretty math
\newcommand{\vect}[1]{\boldsymbol{#1}}

\usepackage[dvipsnames]{xcolor} % colorbox

\title{Speeding up astrochemical reaction networks with autoencoders and neural ODEs}

\author{%
  Immanuel Sulzer\\%\thanks{Use footnote for providing further information about author (webpage, alternative address)} \\
  Interdisciplinary Center\\ for Scientific Computing,\\
  University of Heidelberg,\\
  Im Neuenheimer Feld 205,\\ D-69120 Heidelberg\\
  \texttt{immanuel.sulzer@stud.uni-heidelberg.de} \\
    \And
     Tobias Buck\\ %\thanks{Use footnote for providing further information about author (webpage, alternative address)} \\
  Interdisciplinary Center\\ for Scientific Computing,\\  University of Heidelberg,\\ Im Neuenheimer Feld 205,\\ D-69120 Heidelberg\\
  \texttt{tobias.buck@iwr.uni-heidelberg.de} \\
}

\begin{document}

\maketitle
\begin{abstract} 
In astrophysics, solving complex chemical reaction networks is essential but computationally demanding due to the high dimensionality and stiffness of the ODE systems. Traditional approaches for reducing computational load are often specialized to specific chemical networks and require expert knowledge. This paper introduces a machine learning-based solution employing autoencoders for dimensionality reduction and a latent space neural ODE solver to accelerate astrochemical reaction network computations. Additionally, we propose a cost-effective latent space linear function solver as an alternative to neural ODEs. These methods are assessed on a dataset comprising 29 chemical species and 224 reactions. Our findings demonstrate that the neural ODE achieves a 55x speedup over the baseline model while maintaining significantly higher accuracy by up to two orders of magnitude reduction in relative error. Furthermore, the linear latent model enhances accuracy and achieves a speedup of up to 4000x compared to standard methods.
\end{abstract}

\section{Motivation}
Chemistry plays a pivotal role in astrophysical simulations, influencing phenomena in the interstellar medium \citep{hollenbach_molecule_1979} and star formation processes \citep{glover_star_2007}. The chemical composition of species not only impacts physical processes but also affects observational outcomes, particularly line emissions \citep[e.g.][]{grassi_krome_2014}.
However, solving chemical reaction networks poses challenges, given the need to handle numerous tightly coupled stiff ODEs, arising from distinct reaction timescales. Existing simplification techniques, such as species omission and grouping \citep{nelson_stability_1999}, instantaneous equilibrium assumption \citep{glover_modelling_2010}, or exploiting sparse reaction matrices \citep{grassi_chemical_2013}, often rely heavily on specific chemical networks and may not always yield effective results.
\enlargethispage{\baselineskip}
As an alternative, \citeauthor{grassi_reducing_2022} demonstrated the effectiveness of autoencoders for dimensionality reduction in astrochemistry. Furthermore, the advent of neural ODEs \cite{chen_neural_2019, kidger_neural_2022} has opened new possibilities for novel ODE solver methods in astrochemical modelling. Our approach seeks to integrate these techniques, aiming to create an easily implementable and efficient solver for chemical reaction networks. Our code is publicly available on Github: (URL: \url{https://github.com/Immi000/ChemicalNeuralODE}). The code is documented and can easily adapted for reuse.

%\section{Method}
\section{Latent space neural ODE solver}
\begin{figure}
    \begin{center}
    \includegraphics[width=.6\textwidth]{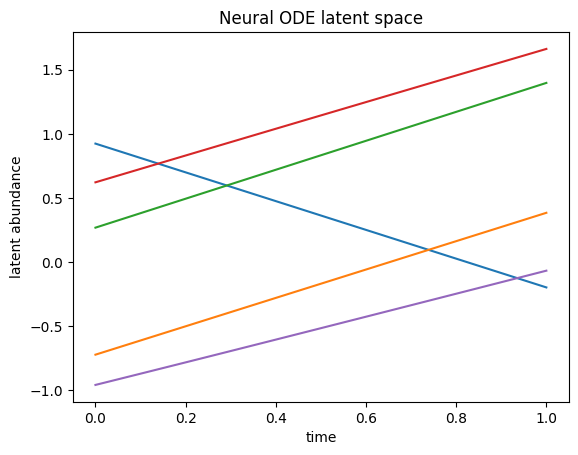}
    \caption{The predicted evolution in the latent space as calculated by the neural ODE (left) and the linear latent model (right). Even though the latent space trajectories of the neural ODE on the left were not constrained to be straight, in some cases the optimal solution found by the optimizer resulted in the simple trajectories shown. This phenomenon inspired the simpler and more performant linear latent model.}
    \label{fig:latent_plots}
    \end{center}
\end{figure}

A neural ODE essentially uses a neural network (NN) $f$ to model the derivative of a vector valued function $\vect{x}(t)$ that depends on a parameter $t$ (time in our application): 
%\begin{equation}
        $\frac{d\vect{x}(t)}{dt} = f(\vect{x}(t), t)$.
%\end{equation}
Our model for solving chemical reaction networks comprises three components: an encoder $\varphi$, the neural ODE $f$, and a decoder $\psi$. The dynamics of the chemical system are characterized by a function $\vect{x}(t)$, where only the initial state $\vect{x}_0$ is known. The encoder transforms the initial state $\vect{x}_0$ into the latent space initial state $\vect{z}_0=\varphi(\vect{x}_0)$, which serves as the initial condition for the ODE:
\begin{minipage}{0.6\textwidth}
%\begin{internallinenumbers}
    %\begin{equation}
        $\frac{d\vect{z}(t)}{dt}=f(\vect{z}(t))$ with $\vect{z}(t=0)=\vect{z}_0$
    %\end{equation}
    where $\vect{z}(t)$ represents the latent space time evolution of the chemical system. $\vect{x}(t)$ can therefore be calculated as:
    \begin{equation}
        \vect{x}(t)=\psi\left(\varphi(\vect{x}_0)+\int_{[t_0,t_1]} f(\vect{z}(t))\,dt\right)
    \end{equation}
    where the integral can be evaluated using any ODE solver. 
    For reasons of performance we have opted for the \verb|odeint_adjoint|-method implemented by \citeauthor{chen_torchdiffeq_2021} with a Dormand-Prince solver \citep{dormand_family_1980}, as it is capable of backpropagation via the adjoint sensitivity method \cite{chen_torchdiffeq_2021, chen_neural_2019}.
    During experimentation, we noticed that some of the trajectories in the latent space are approximately linear (see figure \ref{fig:latent_plots}). This means the latent space ODE is approximately given by a constant $\frac{d\vect{z}(t)}{dt}=f(\vect{z}(t))\approx \vect{c}$. This simplifies integration, obviating the need for a numerical solver or neural ODE. The solution becomes:
$\vect{z}(t) = \int \vect{c}\,dt = \vect{c}t + \vect{z}_0$
enabling direct evaluation at all time points, significantly accelerating overall computation. We evaluate this approach as a "linear latent" model and compare it with the neural ODE approach. As a broader strategy, one can select any analytically integrable function (e.g., a polynomial) to approximate the latent space
%\end{internallinenumbers}
\end{minipage}
\begin{minipage}{0.05\textwidth}
    % spacer
\end{minipage}
\begin{minipage}{0.38\textwidth}
    \includegraphics[width=\textwidth]{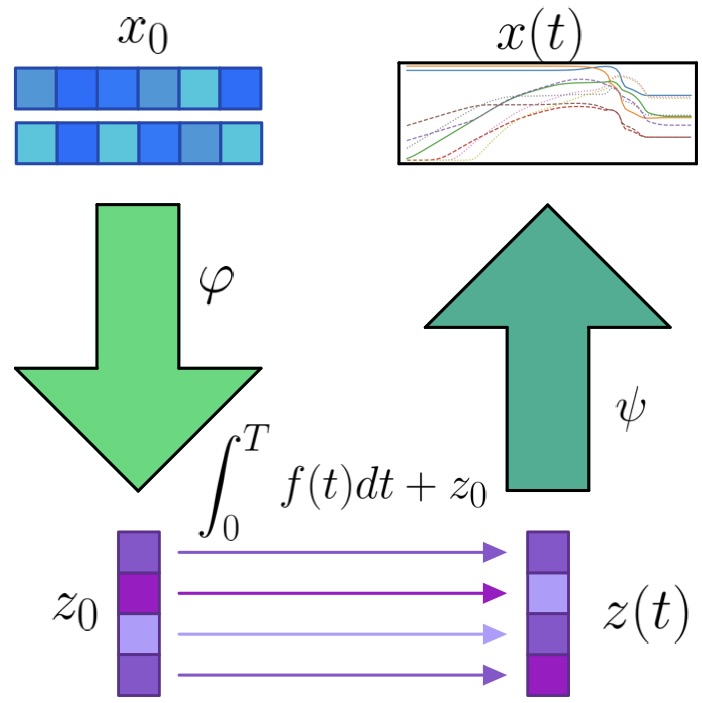}
    \captionof{figure}{
    The encoder $\varphi$ encodes the initial state $\vect{x}_0$ into its latent space equivalent $\vect{z}_0$. The time evolution $\vect{z}(t)$ is then calculated by integrating the neural ODE $f$ with the initial condition $\vect{z}_0$. The calculated latent space evolution of $\vect{z}(t)$ is then transformed back into the real space time evolution of abundances $\vect{x}(t)$ via the decoder $\psi$.}
\end{minipage}
%
%\begin{equation}
%    \frac{d\vect{z}(t)}{dt}=f(\vect{z}(t))\approx \vect{c}
%\end{equation}
%
evolution, preserving the computational advantages of the linear model while enabling higher complexity in the latent space. As the model's ultimate training objective is to discover a latent space representation with greatly simplified dynamics, it draws inspiration from manifold learning and sparse identification of nonlinear dynamics \citep{brunton_discovering_2016}.

\subsection{Training data}
Our training data is based on the chemical network \verb|osu_09_2008| and has been generated following \cite{grassi_reducing_2022}. This network consists of 224 reactions between 29 species %(H, H$^+$, H$_2$, H$_2^+$, H$_3^+$, O, O$^+$, OH$^+$, OH, O$_2$, O$_2^+$, H$_2$O, H$_2$O$^+$, H$_3$O$^+$, C, C$^+$, CH, CH$^+$, CH$_2$, CH$_2^+$, CH$_3$, CH$_3^+$, CH$_4$, CH$_4^+$, CH$5^+$, CO, CO$^+$, HCO$^+$, He, He$^+$, and e$^-$) 
and assumes other physical parameters such as temperature ($T=50$ K), cosmic ray ionization rate ($\zeta=10^{-16}s^{-1}$) and total density ($n_{tot}=10^4$ cm$^{-3}$) to be constant.
We simulate the species evolution 500 times over 100 time steps, varying the initial conditions\footnote{The code for generating the data can be found in the repository of \cite{grassi_reducing_2022}}. The chemical network solves the reactions within a logarithmic space, which is min-max normalized to the range [-1, 1]. The initial state is assumed to be predominantly molecular, with $n_{H_2}$ equal to $n_{\rm tot}$. $n_C$ and $n_{C^+}$ are randomized in log space, ranging between $10^{-6}n_{\rm tot}$ and $10^{-3}n_{\rm tot}$. Molecular Oxygen is initialized to equal $n_O=n_C+n_{C^+}$. The initialization of electrons ensures that the system maintains charge neutrality, while all other species commence at an initial value of $10^{-20}n_{\rm tot}$.
The time evaluation points span an interval of $10^8$ years and are evenly distributed on a logarithmic scale, subsequently min-max normalized to $[0, 1]$.

\subsection{Architecture design and loss function}
Our final model consists of three NNs: the encoder $\varphi$, the decoder $\psi$, and the neural ODE $f$.
The \textbf{encoder $\varphi$} is a fully connected feedforward NN with four hidden layers. The input layer has 29 dimensions, one for each of the 29 chemical species. The four hidden layers have dimensions of 64, 32, 16, and 8, respectively, and the output layer has five dimensions, corresponding to the latent space dimensionality for this experiment. The activation function used in both the encoder and decoder is $\tanh$. 
The \textbf{decoder $\psi$} mirrors the architecture of the encoder in reverse order. 
The \textbf{neural ODE $f$} consists of a fully connected feedforward NN with an input layer, 5 hidden layers of width 64, and an output layer. The activation function after every layer (except the last) is $\tanh$. To allow for a wider range of output gradients, a scaled version of $\tanh$ is used, including a learnable scalar parameter $s$:
\enlargethispage{\baselineskip}
%\begin{equation}
    $\frac{d\vect{z}(t)}{dt} = s \cdot \tanh(\textbf{NN}(\vect{z})/s)$.
%\end{equation}
This modification extends the output of the NN to the range of $[-s, s]$, offering greater flexibility in gradient values and introducing a regularizing effect \cite{tang_reduced_2022}.
For model training, five loss functions are defined:
The \textbf{reconstruction loss ($L_0$)} measures the discrepancy between the predicted trajectory $\vect{x}'(t)$ and the ground truth $\vect{x}(t)$, computed as $L_0 = |\vect{x}(t)-\vect{x}'(t)|^2$.
The \textbf{identity loss ($L_1$)} ensures that the combination of the encoder $\varphi$ and decoder $\psi$ approximates the identity. It is calculated as $L_1 = |\vect{x}(t)-\psi(\varphi(\vect{x}(t)))|^2$.
The \textbf{mass conservation loss ($L_2$)}, incorporating the species masses vector $\vect{m}$, ensures that the mass of the predicted trajectory aligns with the mass of the ground truth at each time point. It is computed as $L_2 = |\vect{m}\cdot \vect{x}(t) - \vect{m}\cdot \vect{x}'(t)|^2$.
\textbf{Gradient losses ($L_3$ and $L_4$)} enforce the predicted trajectories to closely follow the ground truth by comparing first and second-order gradients in real space. $L_3$ measures the discrepancy in first-order gradients, and $L_4$ examines second-order gradients. They are defined as $L_3 = |\frac{d\vect{x}(t)}{dt}-\frac{d\vect{x}'(t)}{dt}|^2$ and $L_4 = |\frac{d^2\vect{x}(t)}{dt^2}-\frac{d^2\vect{x}'(t)}{dt^2}|^2$. After a burn-in phase of ten epochs, the losses are normalized to unity and the $L_0$ loss is weighted by 100. Optimization is performed using the Adam optimizer \citep{kingma_adam_2017} with a cosine-annealing learning rate scheduler, varying the learning rate between $10^{-3}$ and $10^{-5}$.
\section{Results}
\begin{figure}
\begin{center}
    \includegraphics[width=.95\textwidth]{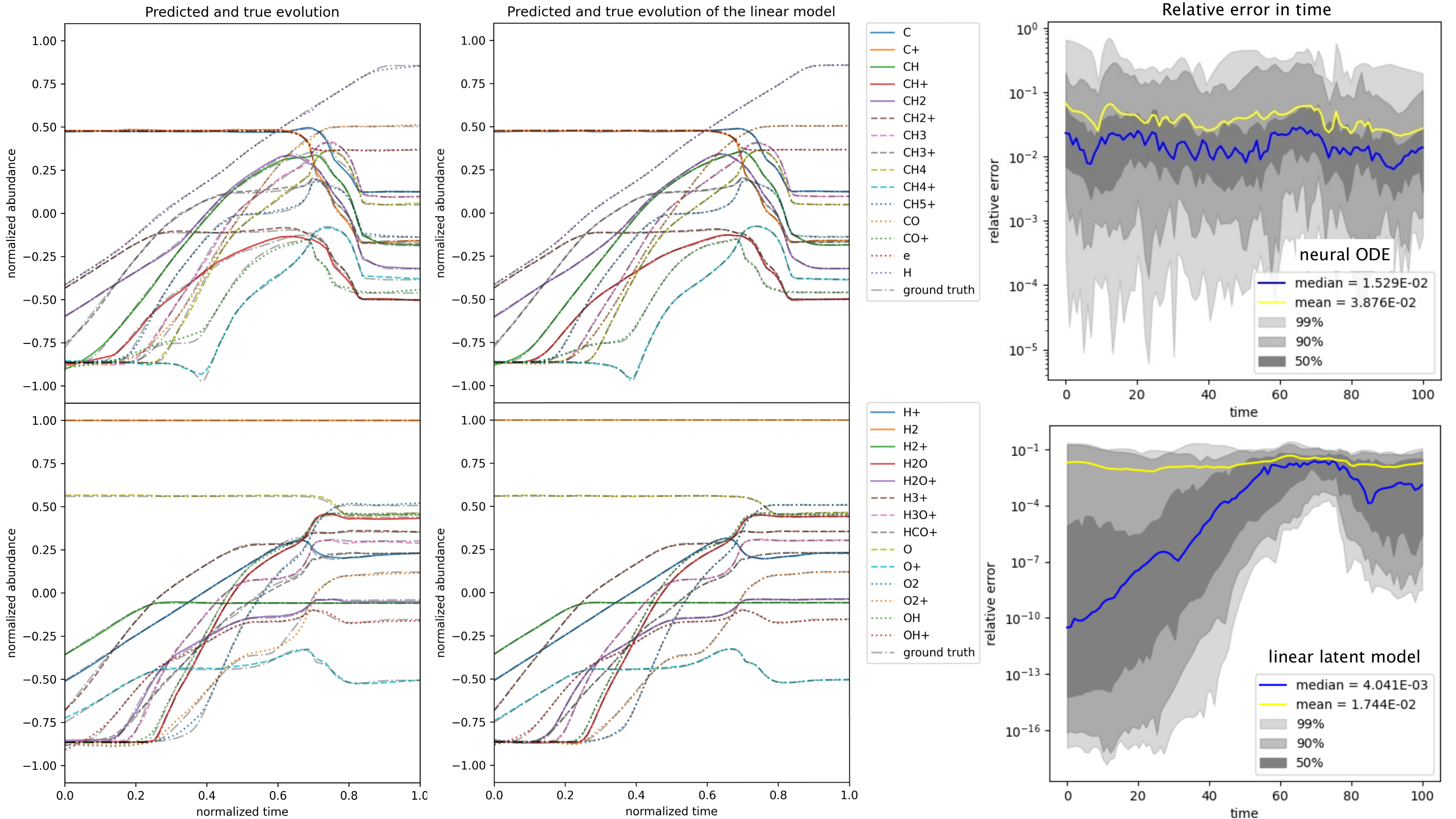}
    \caption{The ground truth (grey) and the predicted abundances (color) from the neural ODE model (left) and the linear latent space model (middle). The right panels show the relative error over time for the neural ODE model (upper panel) and the linear latent model (lower panel).} 
    \label{fig:prediction}
\end{center}
\end{figure}

\enlargethispage{\baselineskip}
The models are evaluated on a test set of 50 unseen trajectories. Fig.\ref{fig:prediction} shows one sample from the test set alongside the predictions from both models: the neural ODE (left panels) and the linear latent model (middle panels). 
The predicted trajectories closely match the ground truth trajectories. For certain species such as H, H$_2$O, or H$_3^+$, there is hardly any discernible deviation while for other species like C or CH$_4$, some minor discrepancies from the ground truth are noticeable. Notably, regions with moderate to strong curvature produce larger errors than flatter trajectories.
Quantitatively, the deviations correspond to a Root Mean Square Error (RMSE) of $9\cdot10^{-3}$ for the neural ODE model and $3\cdot10^{-3}$ for the linear latent model. Additionally, the relative error (rightmost panels in Fig.~\ref{fig:prediction}) for the neural ODE trajectory ranges from $5.96\cdot 10^{-8}$ to $0.58$, with a median relative error of $0.015$. In contrast, the linear latent model exhibits relative errors ranging between $10^{-16}$ and $\sim0.1$, with a median relative error of $0.004$.
In comparison, \citeauthor{grassi_reducing_2022} employed a similar dimensionality reduction approach utilizing autoencoders but modelling reactions in latent space by explicitly defining a sparse reaction matrix. Evaluated on the same dataset, \citeauthor{grassi_reducing_2022} achieve a logarithmic relative error $\langle\log(\Delta x)\rangle$ ranging between $-2.96$ and $0.44$. In contrast, our neural ODE model achieves a log relative error ranging from $-4.41$ to $-1.03$, while our linear latent space model reaches a log relative error ranging from $-5.09$ to $-1.64$. Overall, we have achieved significantly improved reconstruction results, with relative errors several orders of magnitude lower than those reported by \cite{grassi_reducing_2022}. The linear latent space model, in particular, achieves a relative error roughly one order of magnitude lower than that of the neural ODE model. This is further reflected in the logarithmic mean average error across all species, where \citeauthor{grassi_reducing_2022} achieved $-0.41$, while our neural ODE model achieves $-2.40$, and the linear model reports $-2.94$. This represents an improvement of approximately two orders of magnitude compared to previous results.
\subsection{Speed}

Performance measurements of various solver methods were conducted, measuring the time required for a model to predict the evolution of an initial abundance from the test set over 100 time steps. The standard \verb|solve_ivp| integration, which was used with a multi-step variable-order BDF solver and employed in generating the training data, was executed on a Dual Intel Xenon Gold 6254 processor. In contrast, all other methods were run on an Nvidia RTX 2080ti GPU.
In comparison to the reported performance by \citeauthor{grassi_reducing_2022}, who achieved a speed-up factor of 65 over \verb|solve_ivp|, our neural ODE model achieves a speed-up factor of 55. Remarkably, the linear latent space model demonstrates a substantial speed-up, reaching a factor of 4270 significantly outperforming both the neural ODE model and the sparse solver by a factor of approximately 77.
\begin{table}
\caption{Performance measurements of different solver methods.}
    \label{tab:speed}
\begin{center}
    \begin{tabular}{c c c c}
    \toprule
        Model & integration time & faster than \verb|solve_ivp|  & faster than Neural ODE \\
        \hline
        \hline
        scipy \verb|solve_ivp| & $1930$ ms & - & 0.018 \\
        \citeauthor{grassi_reducing_2022} & 29.7 ms & 65 & 1.18 \\
        Neural ODE & $34.9$ ms & 55 & - \\
        Linear Model & $0.45$ ms & 4270 & 77 \\
        \bottomrule
    \end{tabular}
\end{center}
\end{table}

\section{Limitations \& Outlook}

The impressive results of the model should be considered in light of several limitations: The model's performance is constrained by a limited range of initial conditions. %Most chemical species are initialized with small, fixed values, with only C and C$^+$ having truly random initializations. 
%This limitation restricts the diversity of initial conditions and may not fully capture the real-world variability.
The absence of variation in temperature, cosmic ray ionization rate, and total density as initial conditions means that the trajectories cover only a limited portion of the phase space. In reality, these parameters change over time, and incorporating such changes into the chemistry solver is challenging, as suggested by related work \citep{branca_neural_2022}.
The chemical reaction network considered in this study includes 224 reactions and 29 species. However, real-world astrochemical systems can involve much larger networks with over 400 species \citep[e.g.][]{le_teuff_umist_2000}. It remains uncertain whether the methods tested in this study would perform effectively on such extensive networks.
While the overall accuracy is high, there are regions and species, particularly those with low abundances and high reaction rates, where accurate predictions may not be achieved. The dimensionality reduction approach relies on data being projectable to a lower-dimensional manifold while preserving information about reactions. Questions remain about the optimal choice of latent space dimensions and the identification of the latent dynamics.

Looking forward, several promising directions for further research and exploration can be considered:
Investigating the mathematical properties of the autoencoder, particularly why latent space trajectories tend to become straight lines without explicit constraints, could provide valuable insights into the model's behavior.
Further exploration is needed to understand the observed latent space trend where, in five dimensions, two lines typically have negative slopes while the others have positive slopes, albeit with similar magnitudes. 
Exploring advancements in machine learning for modeling physical systems and solving linear equations can enhance the model's capabilities. Approaches such as the LM-architecture proposed by \cite{lu_beyond_2020}, coordinate-based neural operators for solving PDEs \citep{serrano_operator_2023}, and sparse identification methods \citep{brunton_discovering_2016} are promising areas for investigation.
Investigating alternative methods, such as continuous-time echo state networks for modeling disturbed systems \citep{roberts_continuous-time_2022}, can expand the toolbox for solving complex chemical simulations. Exploring these methods may provide complementary or alternative solutions to the challenges posed by large chemical networks.

In summary, while the current model has shown impressive results, there are still opportunities for improvement and avenues for further research to address the limitations and enhance its capabilities in solving complex astrochemical simulations.

\section*{Broader impact statement}

This work purely aims to aid scientific algorithm design to speed-up ODE calculations. Therefore the authors are not aware of any immediate ethical or societal implications of this work.  Thinking more broadly, many relevant systems in a broad range of science can be described by coupled ODEs and as such our approach of accelerating their solution with the help of ML might have beneficial impact beyond astrophysical modelling. This may contribute to a broader application and development of such algorithms and in the long run might help to save energy when modelling complex systems.

\bibliography{bibliography}
\end{document}